# The impact of coercive, normative, and mimetic Stress on Chinese teachers' continuance intention to use generative AI: An integrated perspective of the Expectation-Confirmation Model and Institutional Theory

**First Author**
Kunjie Jia
202400630229@mail.sdu.edu.cn
School of Translation Studies, Shandong University, Weihai, China

**Second Author**
Kai Cui
kcui@springfieldca.org
Springfield Commonwealth Academy, Springfield, MA, USA
https://www.springfieldcommonwealthacademy.org/academics/faculty

**Third Author**
Huimin He
Xi'an Jiaotong-Liverpool University, Suzhou, China
Email: huimin.he@xjtlu.edu.cn

**Fourth and Corresponding Author**
Yiran Du
University of Cambridge, Cambridge, United Kingdom
Email: yd392@cam.ac.uk

**Abstract**
This study investigates Chinese teachers' continuance intention to use generative artificial intelligence (AI) by integrating the Expectation–Confirmation Model with Institutional Theory. A sequential explanatory mixed-methods design was employed. Questionnaire data from 437 teachers were analysed using structural equation modelling, followed by semi-structured interviews with 15 teachers to further interpret the findings. The results indicate that confirmation, perceived usefulness, and satisfaction play important roles in shaping teachers' continuance intention, while institutional pressures, including coercive, normative, and mimetic influences, also contribute to continued use. Qualitative findings further reveal that teachers often use generative AI pragmatically to support tasks such as lesson preparation and idea generation, while simultaneously exercising caution and critically evaluating the reliability of AI-generated content. These findings highlight the combined influence of individual evaluations and institutional contexts on teachers' sustained engagement with generative AI in education.

## 1. Introduction

Generative AI has rapidly become important in education, especially after the release of large language models such as ChatGPT. These tools can support lesson planning, instructional design, explanations, feedback, and other teaching tasks (C. Wang et al., 2026; Lee et al., 2026; Du et al., 2025; Aljohani, 2026). However, concerns remain about reliability, ethics, and teachers' need to critically evaluate AI-generated outputs before classroom use (Weng & Fu, 2025).

Understanding teachers' continued use of generative AI is therefore an important issue in educational technology research. The Expectation–Confirmation Model explains continuance intention through confirmation, perceived usefulness, and satisfaction (Bhattacherjee, 2001), and has been widely applied to educational technologies such as e-learning and AI-supported systems (Alshammari & Alshammari, 2024; Ode et al., 2025; Zheng et al., 2025). However, existing research on generative AI has focused mainly on perceptions, acceptance, or early adoption, with less attention to sustained use in teaching practice.

Prior studies have also tended to emphasise individual evaluations while overlooking the institutional contexts shaping teachers' technology use. Policy directives, organisational expectations, and professional norms can influence teachers' engagement with emerging technologies, particularly in China, where educational digitalisation and AI integration are strongly promoted (Niu et al., 2022; Zhao et al., 2025; Knox, 2023; Yuan, 2024). Drawing on Institutional Theory, this study examines coercive, normative, and mimetic pressures alongside experiential factors to explain Chinese teachers' continuance intention to use generative AI.

## 2. Literature Review

### 2.1 Generative AI in Education

Generative artificial intelligence has rapidly emerged as a transformative technology in education following the public release of large language models such as ChatGPT (C. Wang et al., 2026). These systems can generate human-like text, explanations, feedback, and instructional materials, enabling new forms of AI-assisted teaching and learning (Y. Du, Tang, et al., 2026). In educational contexts, generative AI is increasingly used for tasks such as lesson planning, content generation, assessment design, and personalised feedback (Y. Du et al., 2025). As a result, scholars have begun examining its implications for instructional practices, teacher productivity, and student learning processes (C. Wang et al., 2024).

Recent studies suggest that generative AI has the potential to support teachers' professional work by automating routine tasks and enhancing instructional design (Lee et al., 2026). For example, teachers may use AI tools to draft teaching materials, generate examples for classroom discussion, or provide formative feedback on student work (Aljohani, 2026). At the same time, concerns have been raised regarding issues such as academic integrity, reliability of AI-generated information, and the need for critical AI literacy among educators and students (Weng & Fu, 2025). Consequently, research on generative AI in education has increasingly focused on how teachers evaluate, adopt, and integrate these tools into their pedagogical practices (Y. Wang et al., 2025).

Within the broader field of educational technology research, scholars have frequently used technology adoption models to understand teachers' engagement with new digital tools (Qian, 2025). One widely applied framework is the Expectation-Confirmation Model, which explains post-adoption behaviour by examining how users' confirmation of expectations, perceived usefulness, and satisfaction influence their intention to continue using a technology (Jung & Jo, 2025). While this model has been applied to various educational technologies, the rapid emergence of generative AI introduces new contextual factors, including institutional expectations and professional norms, that may also shape teachers' continued use of these tools (Yu et al., 2024).

### 2.2 The Chinese Context

China provides a distinctive context for examining teachers' adoption of generative AI due to its strong national emphasis on educational digitalisation and artificial intelligence development (Knox, 2023). In recent years, the Chinese government has introduced a series of strategic initiatives promoting AI integration across multiple sectors, including education. Policies related to "smart education" and digital transformation encourage schools to adopt emerging technologies to enhance teaching quality, improve learning efficiency, and modernise educational systems (Yuan, 2024). These policy initiatives create institutional environments that actively encourage experimentation with AI-supported teaching practices (Chai et al., 2024).

Within Chinese schools, teachers operate in organisational structures characterised by strong policy guidance, hierarchical administration, and collective professional cultures (Niu et al., 2022). Educational reforms are often implemented through top-down directives from ministries, local education authorities, and school leadership. Such governance structures may generate institutional pressures that influence teachers' technology adoption decisions (Gao et al., 2026). In particular, teachers may experience expectations to engage with emerging technologies in order to align with policy priorities and demonstrate professional competence in digital teaching practices (Zhao et al., 2025).

At the same time, Chinese teachers also participate in active professional communities where teaching practices are frequently shared and evaluated among peers (Pan & Wang, 2025). Collaborative lesson planning, open classroom observations, and online teacher communities create opportunities for teachers to observe and imitate innovative instructional practices (Kim, 2024). In this environment, the adoption of generative AI tools may be shaped not only by policy directives but also by peer influence and professional norms (J. Chen et al., 2025). Examining teachers' continued use of generative AI in this context therefore provides an opportunity to understand how institutional forces interact with individual experiences in shaping post-adoption technology behaviour (K. Wang et al., 2024).

## 3. Theoretical Framework

This study integrates the Expectation-Confirmation Model and Institutional Theory to examine teachers' continuance intention to use generative AI in the Chinese educational context. The Expectation-Confirmation Model explains users' post-adoption behaviour based on their experiences with a technology (Bhattacherjee, 2001), whereas Institutional Theory emphasises how organisational environments shape individuals' professional practices through social and regulatory pressures (Amenta & Ramsey, 2010). Integrating these perspectives enables the study to capture both teachers' experiential evaluations of generative AI and the institutional influences that may affect their continued use.

According to the Expectation-Confirmation Model, continuance intention is determined by users' post-use evaluations, including confirmation of expectations, perceived usefulness, and satisfaction. When users perceive that a technology meets or exceeds their expectations and provides practical benefits for their tasks, they are more likely to feel satisfied and develop a stronger intention to continue using it. In the context of generative AI in education, Chinese teachers' continued use may therefore depend on whether these tools effectively support teaching activities such as lesson preparation, instructional design, and feedback provision.

**Figure 1. The Conceptual Model**

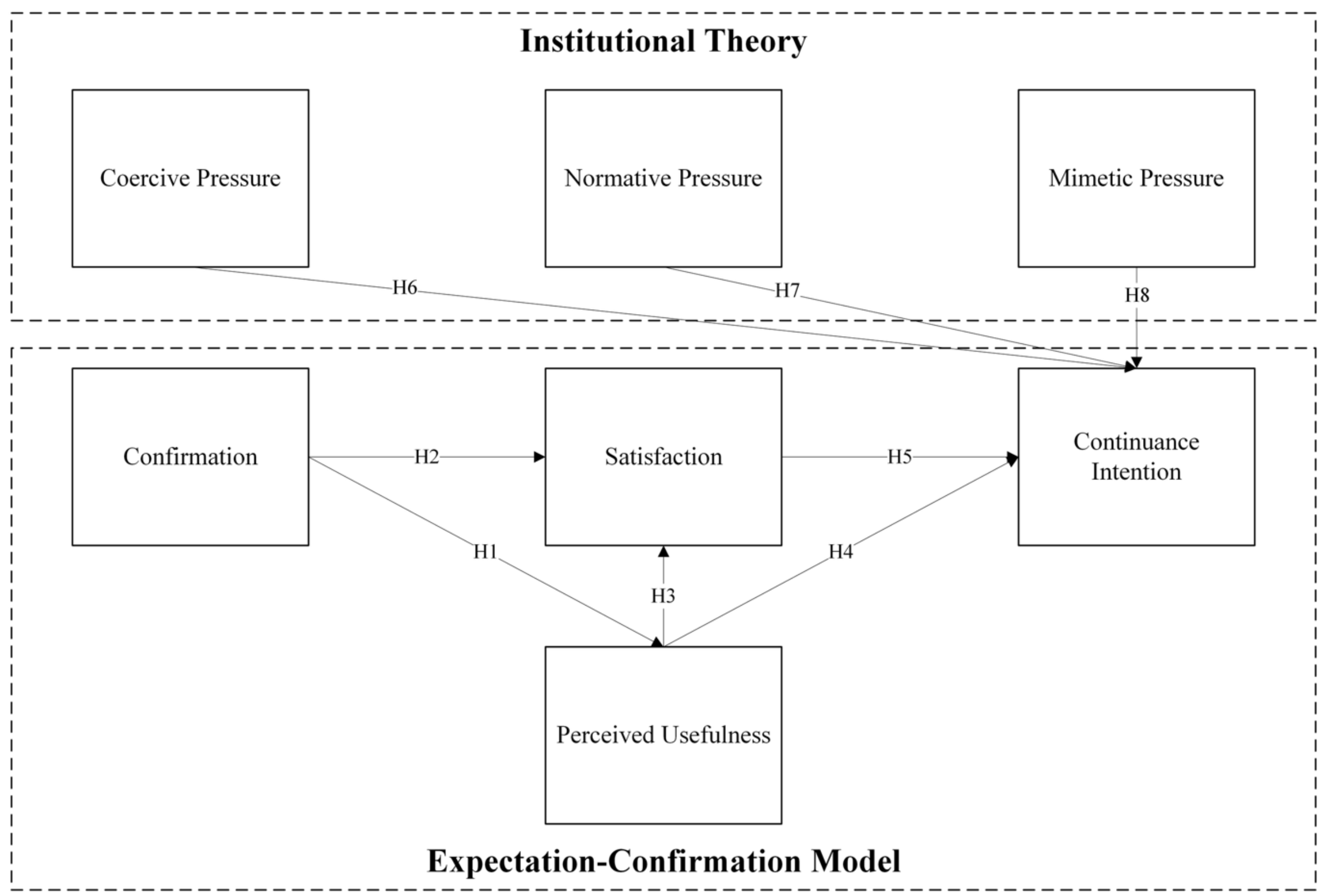

At the same time, teachers' technology use in China is embedded within institutional environments characterised by strong policy guidance and collective professional norms. Institutional Theory suggests that individuals may adopt or continue certain practices due to coercive, normative, and mimetic pressures. In Chinese schools, these pressures may arise from policy initiatives promoting educational digitalisation, expectations within professional communities, and the influence of peer practices. By integrating these perspectives, this study proposes a framework that explains teachers' continued use of generative AI through both experiential factors and institutional influences. The conceptual model and hypothesised relationships are illustrated in Figure 1.

## 4. Research Questions and Hypotheses

### 4.1 The Impact of Confirmation, Perceived Usefulness, and Satisfaction

Based on the conceptual model, the first research question (RQ1) asks: *How do confirmation, perceived usefulness, and satisfaction influence Chinese teachers' continuance intention to use generative AI?* To address this question, this study draws on the Expectation-Confirmation Model, which explains users' post-adoption behaviour based on their evaluations of a technology after initial use (Bhattacherjee, 2001). According to this model, continuance intention is influenced by users' confirmation of expectations, perceived usefulness, and satisfaction. These constructs have been widely applied in studies examining the sustained use of information systems and educational technologies (Ode et al., 2025).

Confirmation reflects the extent to which users' experiences with a technology meet their prior expectations. According to the Expectation-Confirmation Model, when users perceive that a system performs as expected or better, they are more likely to evaluate it as useful and develop positive feelings toward it (C. Zhang et al., 2025). Prior research on information systems continuance shows that confirmation significantly enhances both perceived usefulness and satisfaction (Yu et al., 2024). In the context of generative AI, when Chinese teachers find that these tools effectively support teaching tasks such as lesson preparation or feedback generation, their perceptions of usefulness and satisfaction are likely to increase (Gao et al., 2026).

Perceived usefulness refers to the degree to which users believe that a technology improves their work performance. Teachers who perceive generative AI as beneficial for instructional efficiency or pedagogical support are more likely to feel satisfied with its use (K. Wang et al., 2024). Existing studies indicate that perceived usefulness contributes to satisfaction and directly strengthens continuance intention, while satisfaction serves as a key determinant of sustained system use (Alshammari & Alshammari, 2024; Choi et al., 2023). Therefore, higher perceived usefulness and satisfaction are expected to increase Chinese teachers' continuance intention to use generative AI. Accordingly, the following hypotheses are proposed.

H1: Confirmation positively influences perceived usefulness.
H2: Confirmation positively influences satisfaction.
H3: Perceived usefulness positively influences satisfaction.
H4: Perceived usefulness positively influences continuance intention.
H5: Satisfaction positively influences continuance intention.

### 4.2 The Impact of Coercive Pressure, Normative Pressure, and Mimetic Pressure

Based on the conceptual model, the second research question (RQ2) asks: *How do coercive pressure, normative pressure, and mimetic pressure influence Chinese teachers' continuance intention to use generative AI?* To address this question, this study draws on Institutional Theory, which suggests that individuals' practices are shaped by institutional environments through various forms of social and organisational pressure. In educational settings, teachers' technology use may therefore be influenced not only by their personal evaluations of a technology but also by institutional expectations within schools and the broader educational system.

Institutional theory proposes that institutional isomorphism occurs through three primary mechanisms: coercive, normative, and mimetic pressures (Zhuang & Sun, 2023). Coercive pressure arises from formal regulations, policies, or administrative expectations that encourage or require the adoption of particular practices (Lu & Wang, 2023). In the Chinese educational context, national and local policies promoting digital transformation and artificial intelligence in education may create expectations for teachers to engage with emerging technologies such as generative AI (Knox, 2023). These institutional directives may motivate teachers to continue using AI tools in order to align with policy priorities and organisational expectations (Henadirage & Gunarathne, 2023).

Normative pressure originates from professional norms, training programmes, and expectations within professional communities (Hom et al., 2025). Teachers often operate within collaborative professional cultures where instructional practices are shared and evaluated among colleagues (Chounta et al., 2022). Mimetic pressure occurs when individuals imitate practices perceived as effective or legitimate, particularly when dealing with new and uncertain technologies (Latif et al., 2020). In the case of generative AI, teachers may be influenced by colleagues, peer schools, or professional networks that demonstrate successful use of AI-supported teaching practices. Accordingly, the following hypotheses are proposed.

H6: Coercive pressure positively influences continuance intention.
H7: Normative pressure positively influences continuance intention.
H8: Mimetic pressure positively influences continuance intention.

## 5. Methods

### 5.1 Research Design

This study used a sequential explanatory mixed-methods design, with quantitative data collection and analysis followed by qualitative inquiry. First, a questionnaire survey examined relationships among constructs in the conceptual model, including Expectation–Confirmation Model factors—confirmation, perceived usefulness, and satisfaction—and institutional pressures, namely coercive, normative, and mimetic influences. Second, semi-structured interviews were conducted with purposively selected participants to explore teachers' experiences of continued generative AI use. The qualitative findings helped interpret and contextualise the quantitative results, providing a fuller understanding of factors shaping teachers' continuance intention.

### 5.2 Participants

A total of 437 Chinese teachers completed the questionnaire, recruited online through professional and institutional channels across primary, secondary, and tertiary education. Participation was voluntary and anonymous, with no identifying information collected. Participant demographics are reported in Appendix A. Respondents included 249 female teachers (57.0%) and 188 male teachers (43.0%). Most were aged 25–34 (43.0%), followed by those under 25 (23.3%), aged 35–44 (22.0%), and aged 45 or above (11.7%). Participants taught both STEM (46.9%) and non-STEM subjects (53.1%), with varied teaching experience and educational levels.

For the qualitative phase, 15 interviewees were purposively selected from questionnaire respondents who agreed to follow-up contact. Selection aimed to include teachers with high and low AI continuance intention, based on top- and bottom-quartile scores on the continuance intention scale, while also ensuring diversity in age, gender, subject, experience, and teaching level. Interview demographics are summarised in Appendix B. Interviews continued until thematic saturation was reached.

The study followed standard ethical procedures. Participation was voluntary, informed consent was obtained, and participants were informed of the study purpose, withdrawal rights, and confidentiality measures. Questionnaire data were anonymous, and interview data were anonymised using participant codes.

### 5.3 Questionnaires

Data were collected using a structured questionnaire designed to measure the constructs in the proposed conceptual model. The questionnaire included seven constructs: confirmation (Zheng et al., 2025), perceived usefulness (Chai et al., 2024), satisfaction (Sharma et al., 2025), and continuance intention (Obeid et al., 2024) derived from the Expectation-Confirmation Model, and coercive pressure (Sukoco et al., 2022), normative pressure (Hom et al., 2025), and mimetic pressure (Latif et al., 2020) derived from Institutional Theory. Measurement items were adapted from established scales in prior research and slightly modified to fit the context of teachers' use of generative AI in teaching (Y. Du, 2024). The questionnaire was administered in Chinese to ensure participants' comprehension, and a translation and back-translation procedure was conducted to maintain equivalence between the English and Chinese versions (Brislin, 1970). A pilot study with 30 teachers was conducted to assess item clarity, leading to minor wording revisions. The questionnaire also collected demographic information and invited respondents to indicate their willingness to participate in a follow-up interview. All items were measured using a seven-point Likert scale ranging from 1 (strongly disagree) to 5 (strongly agree). The full list of constructs and measurement items is presented in Appendix C.

### 5.4 Semi-Structured Interviews

To complement the questionnaire results, semi-structured interviews were conducted to obtain deeper insights into teachers' experiences and perspectives regarding the continued use of generative AI in teaching. The interview protocol was developed based on the hypotheses in the proposed conceptual model, with questions designed to explore how factors such as confirmation, perceived usefulness, satisfaction, and institutional pressures (coercive, normative, and mimetic) shape teachers' intentions to continue using generative AI. Semi-structured interviews were employed to provide participants with the flexibility to elaborate on their experiences while ensuring that key themes related to the research questions were consistently addressed. Teachers who indicated their willingness to participate in a follow-up interview in the questionnaire were subsequently contacted and invited to take part. The interviews were conducted individually in Chinese and lasted approximately 30–45 minutes. After informed consent was obtained, the interviews were carried out following the interview protocol, with additional probing questions used when necessary to clarify responses and encourage further discussion. The complete interview protocol is provided in Appendix D.

### 5.5 Data Analysis

Quantitative data were analysed in R using a two-stage SEM approach (Kline, 2023). Preliminary screening examined missing data, outliers, variable distributions, skewness, and kurtosis. CFA was then conducted to evaluate the measurement model, including model fit, reliability, convergent validity, and discriminant validity. Internal consistency was assessed using Cronbach's α and composite reliability, convergent validity through standardised factor loadings and AVE, and discriminant validity using the Fornell–Larcker criterion. After confirming satisfactory measurement properties, the structural model was estimated to test the hypothesised relationships.

Qualitative interview data were analysed through thematic analysis using a combined deductive–inductive strategy (Tisdell, 2025). An initial coding framework was developed from the conceptual model, while allowing new themes to emerge from the data. Two researchers independently coded a subset of transcripts, with substantial inter-coder reliability indicated by Cohen's kappa ($\kappa = 0.86$). Discrepancies were resolved through discussion, and the refined coding scheme was applied to the remaining transcripts. Final codes were organised into themes reflecting teachers' evaluations, experiences, and continuance intentions regarding generative AI use.

## 6. Results

### 6.1 Quantitative Results

Descriptive statistics for the study constructs are presented in Appendix E. The mean scores of the constructs ranged from 3.21 to 3.92 on a five-point Likert scale, with standard deviations ranging from 0.68 to 0.83. Skewness values ranged from −0.56 to −0.15 and kurtosis values ranged from −0.36 to 0.24. These values fall within the commonly accepted thresholds for normality ($|\text{skewness}| < 2$; $|\text{kurtosis}| < 7$), indicating that the data distribution did not substantially deviate from normality and was appropriate for subsequent structural equation modelling.

**Figure 2. Structural Model Results**

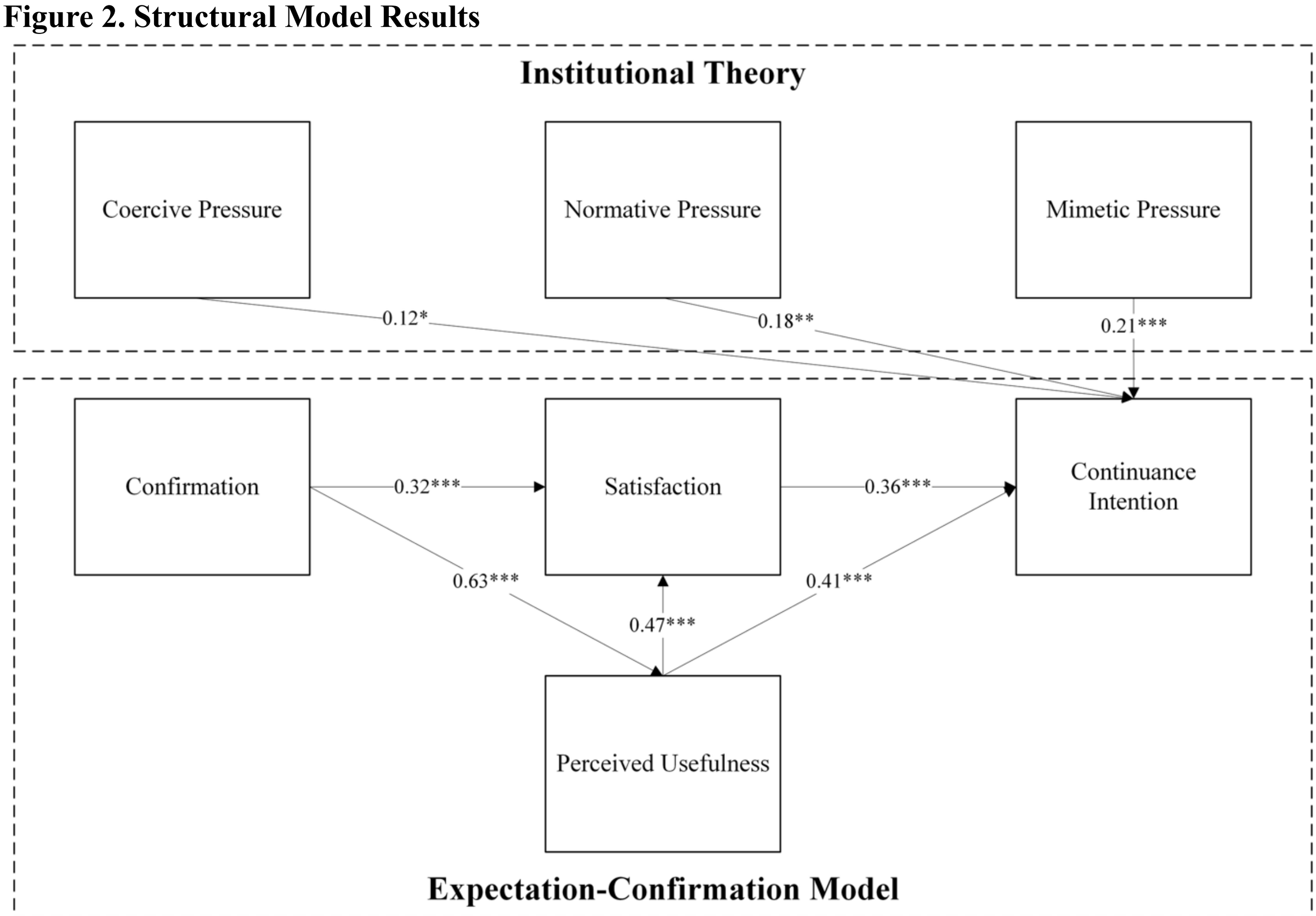

Confirmatory factor analysis was conducted to assess the measurement model. As shown in Appendix F, the measurement model demonstrated satisfactory model fit ($\chi^2/df = 2.21$, CFI = 0.95, TLI = 0.94, RMSEA = 0.053, SRMR = 0.046), meeting recommended criteria. The results for reliability and convergent validity are presented in Appendix G. All factor loadings exceeded 0.70, Cronbach's α values ranged from 0.83 to 0.90, and composite reliability values ranged from 0.86 to 0.91, indicating acceptable internal consistency. In addition, the average variance extracted (AVE) values for all constructs exceeded the recommended threshold of 0.50, supporting convergent validity. Discriminant validity was examined using the Fornell–Larcker criterion (Appendix H), and the square roots of AVE for each construct were greater than the correlations with other constructs.

After establishing the adequacy of the measurement model, the structural model was evaluated to test the proposed hypotheses. The structural model also demonstrated acceptable fit ($\chi^2/df = 2.34$, CFI = 0.94, TLI = 0.93, RMSEA = 0.057, SRMR = 0.049) (Appendix F). The results of the structural model analysis are presented in Appendix I and Figure 2. Confirmation had significant positive effects on perceived usefulness ($\beta = 0.63, p < .001$) and satisfaction ($\beta = 0.32, p < .001$), and perceived usefulness significantly influenced satisfaction ($\beta = 0.47, p < .001$). Both perceived usefulness ($\beta = 0.41, p < .001$) and satisfaction ($\beta = 0.36, p < .001$) had significant positive effects on continuance intention. In addition, coercive pressure ($\beta = 0.12, p < .05$), normative pressure ($\beta = 0.18, p < .01$), and mimetic pressure ($\beta = 0.21, p < .001$) were all significantly associated with continuance intention. All hypothesised relationships were supported.

### 6.2 Qualitative Results

The qualitative analysis drew on interviews with fifteen teachers representing both high and low continuance intention groups and diverse demographic and professional backgrounds (Appendix B). Participants differed in age, subject area, teaching level, and experience, allowing the analysis to capture a range of perspectives regarding the use of generative AI in teaching. Three overarching themes emerged from the thematic analysis: perceived practical usefulness of generative AI, critical evaluation and cautious trust, and institutional and peer influences on technology use. These themes provide a deeper explanation of the patterns observed in the quantitative results.

The first theme concerns teachers' perceptions of practical usefulness and efficiency gains. Many participants in the high continuance intention group described generative AI as a tool that could reduce routine workload and support instructional preparation. For example, P4 explained that "when preparing lessons, the AI can quickly generate examples or explanations, which helps me organise the material faster." Similarly, P7 noted that AI tools were helpful for brainstorming instructional ideas when designing classroom activities. However, even among teachers who viewed AI positively, usefulness was often framed in pragmatic rather than transformative terms. Several participants emphasised that AI outputs usually required further editing before classroom use. As P5 observed, "the AI can give a basic structure, but teachers still need to revise it carefully." This suggests that perceived usefulness may be linked primarily to efficiency improvements rather than full pedagogical reliance on AI-generated materials.

The second theme reflects teachers' cautious trust and critical evaluation of AI-generated content. Participants from both continuance intention groups reported concerns regarding the reliability and accuracy of generative AI outputs. Teachers frequently described the need to verify information before incorporating it into teaching materials. P10 commented that "sometimes the AI provides answers that look reasonable but are not completely accurate." Similarly, P11 stated that she used AI mainly "to get ideas, but I always check the information myself." Such accounts indicate that teachers often approached AI as a supportive ideation tool rather than an authoritative knowledge source. Notably, participants with low continuance intention tended to emphasise these limitations more strongly, sometimes describing the verification process as time-consuming. For instance, P13 argued that "if I need to check everything again, it may not actually save time."

The third theme concerns the influence of institutional expectations and professional norms on teachers' engagement with generative AI. Several participants referred to policy initiatives and school-level encouragement to explore digital technologies. For example, P2 explained that "the school promotes the use of new technologies, so many teachers are encouraged to experiment with AI tools." At the same time, peer influence also played an important role. Teachers often reported learning about generative AI through colleagues or professional development activities. P1 noted that "when other teachers share how they use AI, it makes you curious to try it as well." Nevertheless, some teachers indicated that such pressures could lead to symbolic or exploratory use rather than sustained integration. As P12 commented, "sometimes teachers try AI because it is fashionable, but not everyone continues to use it regularly."

Overall, the qualitative findings suggest that teachers' continuance intention is shaped by a complex interplay between perceived efficiency benefits, critical evaluation of AI reliability, and institutional pressures encouraging experimentation. While many participants recognised the potential value of generative AI for supporting teaching tasks, their continued use was often accompanied by caution and selective adoption. These insights help contextualise the quantitative findings by illustrating how teachers balance perceived usefulness with concerns about reliability and professional responsibility when deciding whether to integrate generative AI into their ongoing teaching practices.

## 7. Discussion

### 7.1 The Impact of Confirmation, Perceived Usefulness, and Satisfaction

The findings reinforce the Expectation–Confirmation Model by showing that confirmation, perceived usefulness, and satisfaction are key determinants of teachers' continuance intention to use generative AI. Consistent with prior continuance research (Bhattacherjee, 2001), confirmation shaped teachers' post-adoption evaluations by influencing whether they perceived generative AI as useful after initial use. When AI tools met or exceeded expectations, teachers were more likely to recognise their value for instructional tasks, echoing previous findings on AI-based educational technologies (Yu et al., 2024; Zhang et al., 2025).

However, the qualitative findings suggest that this confirmation was largely pragmatic. Teachers valued generative AI mainly for routine instructional preparation and idea generation, rather than as a tool for transforming pedagogy. Thus, its perceived usefulness currently appears to relate more to efficiency than to deeper changes in teaching practice.

The results also highlight the roles of perceived usefulness and satisfaction in sustaining AI use. Teachers who saw clear professional benefits were more likely to report satisfaction and continued engagement (Alshammari & Alshammari, 2024; Choi et al., 2023; K. Wang et al., 2024). However, satisfaction was often accompanied by caution, as teachers stressed the need to verify AI-generated outputs due to concerns about accuracy and reliability. This suggests that continued use reflects selective, instrumental adoption rather than uncritical reliance on AI-generated instructional support.

### 7.2 The Impact of Coercive Pressure, Normative Pressure, and Mimetic Pressure

The findings also highlight the role of institutional pressures in shaping teachers' continuance intention to use generative AI. Drawing on Institutional Theory, the study shows that coercive, normative, and mimetic pressures influence teachers' engagement with emerging technologies. In China, coercive pressure often stems from policy initiatives and administrative expectations promoting digital transformation and AI integration (Knox, 2023). Such pressures may encourage teachers to experiment with AI tools to align with institutional priorities and demonstrate professional competence, consistent with prior research on organisational influences in technology adoption (Lu & Wang, 2023).

However, the qualitative findings suggest that institutional pressure does not always lead to deep pedagogical integration. Some teachers used AI mainly because it was promoted by schools or policy initiatives, indicating that institutional encouragement may produce exploratory or symbolic use rather than sustained instructional transformation.

Normative and mimetic pressures also shaped AI use. Professional norms and peer exchanges encouraged teachers to view AI adoption as part of digital competence and innovation (Chounta et al., 2022), while observing colleagues' AI use stimulated curiosity and experimentation (Latif et al., 2020). Yet imitation alone did not ensure continued adoption. Teachers were likely to sustain AI use only when they perceived clear practical value and considered the verification effort manageable. Thus, institutional pressures may promote initial engagement, but long-term continuance depends on teachers' evaluations of usefulness and reliability.

### 7.3 Theoretical and Practical Implications

This study makes several theoretical contributions to research on educational technology adoption and generative AI in education. By integrating the Expectation–Confirmation Model with Institutional Theory, it extends continuance intention research beyond individual-level factors such as perceived usefulness and satisfaction (Jung & Jo, 2025; Zheng et al., 2025). The findings show that teachers' continued use of generative AI is shaped by both experiential evaluations and institutional influences, including policy expectations and professional norms. This offers a broader account of technology use as embedded in organisational and policy contexts.

Practically, the findings suggest that sustained AI use depends less on institutional promotion alone and more on teachers' perceived practical value and confidence in evaluating AI-generated outputs. Professional development should therefore prioritise critical AI literacy, including the ability to assess the accuracy, limitations, and pedagogical relevance of AI-generated content. Schools should also support collaborative professional communities where teachers can share strategies and reflect on AI integration, helping move from exploratory use to more purposeful pedagogical practice.

### 7.4 Limitations and Future Directions

This study has several limitations. First, it relied mainly on self-reported questionnaire data, which may be affected by common method bias and social desirability (X. Chen et al., 2022; C. Du et al., 2025; Y. Du, 2023, 2024, 2025b, 2025a, 2026; Y. Du et al., 2024, 2025; Y. Du, Li, et al., 2026; Y. Du, Tang, et al., 2026; Y. Du, Yuan, et al., 2026; Y. Du & He, 2026e, 2026c, 2026d, 2026b, 2026a; He & Du, 2024; Tang, Jia, et al., 2026; Tang, Lau, et al., 2026; C. Wang et al., 2024, 2026; W. Zhang et al., 2026; Zou et al., 2023, 2024). Second, the sample included only teachers in China, so the findings may not generalise to other educational systems with different policy, institutional, and professional contexts. Third, although the mixed-methods design captured teachers' perspectives, it did not include direct classroom observation; therefore, the findings reflect perceived rather than observed AI use.

Future research should use more diverse methods and contexts. Cross-national or cross-institutional studies could examine how policy and professional cultures shape teachers' continuance intention to use generative AI. Longitudinal designs could track how perceptions and usage patterns change over time. Classroom observations and learning analytics could also provide stronger evidence of how generative AI is actually integrated into teaching and learning, and how it affects instructional practice and student outcomes.

## 8. Conclusion

This study examined Chinese teachers' continuance intention to use generative AI by integrating the Expectation–Confirmation Model with Institutional Theory. The findings show that continued use is shaped by both experiential evaluations and institutional influences. Confirmation, perceived usefulness, and satisfaction were important predictors, suggesting that teachers are more likely to continue using AI when they perceive practical teaching benefits. Institutional pressures, including coercive, normative, and mimetic influences, also shaped AI use within the wider educational context. Qualitative findings further indicate that teachers used generative AI cautiously, mainly as a supportive tool for idea generation and instructional preparation. Overall, the study offers a more comprehensive account of teachers' continued AI use and provides implications for responsible AI integration in education.

## Appendices

### Appendix A. Questionnaire Participant Characteristics (*N* = 437)

| Variable | Category | Frequency (*n*) | Percentage (%) |
|---|---|---|---|
| Age | Under 25 | 102 | 23.3 |
| | 25–34 | 188 | 43.0 |
| | 35–44 | 96 | 22.0 |
| | 45 or above | 51 | 11.7 |
| Gender | Male | 188 | 43.0 |
| | Female | 249 | 57.0 |
| Teaching subject | STEM | 205 | 46.9 |
| | Non-STEM | 232 | 53.1 |
| Teaching experience | Less than 3 years | 120 | 27.5 |
| | 3–5 years | 109 | 24.9 |
| | 6–10 years | 116 | 26.5 |
| | More than 10 years | 92 | 21.1 |
| Teaching level | Primary | 154 | 35.2 |
| | Secondary | 150 | 34.3 |
| | Tertiary | 133 | 30.5 |

### Appendix B. Interview Participant Characteristics (*N* = 15)

| ID | Continuance Intention | Age | Gender | Teaching Subject | Teaching Experience | Teaching Level |
|---|---|---|---|---|---|---|
| P1 | High | 25–34 | Female | STEM | 3–5 years | Secondary |
| P2 | High | 35–44 | Male | Non-STEM | 6–10 years | Secondary |
| P3 | High | Under 25 | Female | STEM | Less than 3 years | Primary |
| P4 | High | 25–34 | Male | STEM | 3–5 years | Secondary |
| P5 | High | 35–44 | Female | Non-STEM | More than 10 years | Tertiary |
| P6 | High | 45 or above | Male | STEM | More than 10 years | Tertiary |
| P7 | High | 25–34 | Female | Non-STEM | 6–10 years | Primary |
| P8 | Low | 25–34 | Male | STEM | 3–5 years | Secondary |
| P9 | Low | Under 25 | Female | Non-STEM | Less than 3 years | Primary |
| P10 | Low | 35–44 | Male | STEM | 6–10 years | Secondary |
| P11 | Low | 25–34 | Female | Non-STEM | 3–5 years | Secondary |
| P12 | Low | 45 or above | Male | Non-STEM | More than 10 years | Tertiary |
| P13 | Low | 35–44 | Female | STEM | 6–10 years | Secondary |
| P14 | Low | 25–34 | Male | Non-STEM | 3–5 years | Primary |
| P15 | Low | Under 25 | Female | STEM | Less than 3 years | Secondary |

**Appendix C. Constructs and Measurement Items**

| Construct | Item | Measurement Item (English) | Measurement Item (Chinese) |
|---|---|---|---|
| Confirmation (CON) (Zheng et al., 2025) | CON1 | My experience with generative AI in teaching has met my expectations. | 我在教学中使用生成式人工智能的体验符合我的预期。 |
| | CON2 | The performance of generative AI in teaching is consistent with what I expected. | 生成式人工智能在教学中的表现与我原本的预期一致。 |
| | CON3 | Overall, my experience with generative AI has confirmed my expectations about its use in teaching. | 总体而言，我使用生成式人工智能进行教学的体验与我的预期相符。 |
| Perceived Usefulness (PU) (Chai et al., 2024) | PU1 | Using generative AI improves my effectiveness in teaching tasks. | 使用生成式人工智能能够提高我完成教学任务的效果。 |
| | PU2 | Generative AI helps me accomplish teaching tasks more efficiently. | 生成式人工智能能够帮助我更高效地完成教学任务。 |
| | PU3 | Generative AI enhances the overall quality of my teaching activities. | 生成式人工智能有助于提升我的整体教学质量。 |
| Satisfaction (SAT) (Sharma et al., 2025) | SAT1 | I am satisfied with my experience of using generative AI in teaching. | 我对在教学中使用生成式人工智能的体验感到满意。 |
| | SAT2 | My experience of using generative AI in teaching has been positive. | 我在教学中使用生成式人工智能的整体体验是积极的。 |
| | SAT3 | Overall, I am pleased with the results obtained from using generative AI in teaching. | 总体而言，我对使用生成式人工智能辅助教学所取得的效果感到满意。 |
| Continuance Intention (CI) (Obeid et al., 2024) | CI1 | I intend to continue using generative AI in my teaching in the future. | 未来我打算继续在教学中使用生成式人工智能。 |
| | CI2 | I expect to use generative AI regularly in my teaching activities. | 我预计会在教学活动中持续使用生成式人工智能。 |
| | CI3 | I plan to increase my use of generative AI in teaching. | 我计划在教学中更多地使用生成式人工智能。 |
| Coercive Pressure (CP) (Hom et al., 2025) | CP1 | My institution encourages teachers to use generative AI in teaching. | 我所在的学校或机构鼓励教师在教学中使用生成式人工智能。 |
| | CP2 | Institutional policies promote the use of generative AI in teaching. | 我所在机构的相关政策推动教师在教学中使用生成式人工智能。 |
| | CP3 | I feel that my institution expects teachers to use generative AI in teaching. | 我感到所在机构期望教师在教学中使用生成式人工智能。 |
| Normative Pressure (NP) (Hom et al., 2025) | NP1 | Teachers in my professional community believe that generative AI should be used in teaching. | 我所在的教师群体普遍认为教师应该在教学中使用生成式人工智能。 |
| | NP2 | Professional norms encourage teachers to adopt generative AI in teaching. | 教师职业发展的相关规范鼓励在教学中使用生成式人工智能。 |
| | NP3 | Professional training and development activities emphasise the use of generative AI in teaching. | 教师培训和专业发展活动强调在教学中使用生成式人工智能。 |
| Mimetic Pressure (MP) (Latif et al., 2020) | MP1 | I am more likely to use generative AI when I see other teachers using it. | 当看到其他教师使用生成式人工智能时，我也更可能去使用它。 |
| | MP2 | Observing other teachers successfully use generative AI encourages me to use it. | 看到其他教师成功使用生成式人工智能会促使我也尝试使用。 |
| | MP3 | I tend to follow teaching practices involving generative AI that are used by other teachers. | 我倾向于借鉴其他教师在教学中使用生成式人工智能的做法。 |

**Appendix D. Semi-Structured Interview Protocol**

| Hypothesis | Interview Question (English) | Interview Question (Chinese) |
|---|---|---|
| H1: Confirmation → Perceived Usefulness | When your experience using generative AI in teaching matched or exceeded your expectations, how did this influence your perception of its usefulness? | 当你在教学中使用生成式人工智能的体验符合或超过预期时，这对你如何看待它的有用性产生了什么影响？ |
| H2: Confirmation → Satisfaction | In what ways did your expectations about generative AI shape your level of satisfaction after using it in teaching? | 你最初对生成式人工智能的预期在多大程度上影响了你使用后的满意度？ |
| H3: Perceived Usefulness → Satisfaction | How does the usefulness of generative AI in your teaching affect how satisfied you feel about using it? | 生成式人工智能在教学中的有用程度如何影响你对使用它的满意度？ |
| H4: Perceived Usefulness → Continuance Intention | To what extent does the usefulness of generative AI influence your decision to continue using it in teaching? | 生成式人工智能在教学中的实用性在多大程度上影响你是否愿意继续使用它？ |
| H5: Satisfaction → Continuance Intention | How does your overall satisfaction with generative AI affect your intention to keep using it in your teaching? | 你对生成式人工智能的整体满意度如何影响你继续在教学中使用它的意愿？ |
| H6: Coercive Pressure → Continuance Intention | How do institutional policies or expectations from your school influence your decision to continue using generative AI in teaching? | 学校或相关机构的政策和要求如何影响你继续在教学中使用生成式人工智能的决定？ |
| H7: Normative Pressure → Continuance Intention | How do professional norms, training programmes, or expectations within the teaching community influence your continued use of generative AI? | 教师群体中的职业规范、培训或专业发展要求如何影响你继续使用生成式人工智能？ |

| H8: Mimetic Pressure → Continuance Intention | How does observing other teachers using generative AI influence your own decision to continue using it in teaching? | 看到其他教师使用生成式人工智能会如何影响你自己继续在教学中使用它的决定？ |
|---|---|---|

**Appendix E. Descriptive Statistics of the Constructs**

| Construct | *M* | SD | Skewness | Kurtosis |
|---|---|---|---|---|
| Confirmation (CON) | 3.78 | 0.71 | -0.48 | 0.12 |
| Perceived Usefulness (PU) | 3.92 | 0.68 | -0.56 | 0.24 |
| Satisfaction (SAT) | 3.65 | 0.74 | -0.39 | -0.05 |
| Continuance Intention (CI) | 3.84 | 0.72 | -0.52 | 0.18 |
| Coercive Pressure (CP) | 3.21 | 0.83 | -0.15 | -0.36 |
| Normative Pressure (NP) | 3.44 | 0.79 | -0.27 | -0.21 |
| Mimetic Pressure (MP) | 3.37 | 0.81 | -0.23 | -0.29 |

**Appendix F. Model Fit Indices**

| Fit Index | Threshold | Measurement Model | Structural Model |
|---|---|---|---|
| $\chi^2/df$ | < 3.00 | 2.21 | 2.34 |
| CFI | ≥ 0.90 | 0.95 | 0.94 |
| TLI | ≥ 0.90 | 0.94 | 0.93 |
| RMSEA | ≤ 0.08 | 0.053 | 0.057 |
| SRMR | ≤ 0.08 | 0.046 | 0.049 |

**Appendix G. Reliability and Convergent Validity**

| Construct | Item | Factor Loading | Cronbach's α | CR | AVE |
|---|---|---|---|---|---|
| Confirmation (CON) | CON1 | 0.78 | 0.86 | 0.88 | 0.70 |
| | CON2 | 0.84 | | | |
| | CON3 | 0.87 | | | |
| Perceived Usefulness (PU) | PU1 | 0.83 | 0.89 | 0.90 | 0.74 |
| | PU2 | 0.88 | | | |
| | PU3 | 0.86 | | | |
| Satisfaction (SAT) | SAT1 | 0.81 | 0.87 | 0.89 | 0.72 |
| | SAT2 | 0.85 | | | |
| | SAT3 | 0.87 | | | |
| Continuance Intention (CI) | CI1 | 0.86 | 0.90 | 0.91 | 0.78 |
| | CI2 | 0.89 | | | |
| | CI3 | 0.88 | | | |
| Coercive Pressure (CP) | CP1 | 0.76 | 0.84 | 0.86 | 0.67 |
| | CP2 | 0.82 | | | |
| | CP3 | 0.85 | | | |
| Normative Pressure (NP) | NP1 | 0.79 | 0.85 | 0.87 | 0.69 |
| | NP2 | 0.84 | | | |
| | NP3 | 0.83 | | | |
| Mimetic Pressure (MP) | MP1 | 0.77 | 0.83 | 0.86 | 0.66 |
| | MP2 | 0.82 | | | |
| | MP3 | 0.84 | | | |

**Appendix H. Discriminant Validity (Fornell–Larcker Criterion)**

| Construct | CON | PU | SAT | CI | CP | NP | MP |
|---|---|---|---|---|---|---|---|
| Confirmation (CON) | **0.84** | | | | | | |
| Perceived Usefulness (PU) | 0.63 | **0.86** | | | | | |
| Satisfaction (SAT) | 0.59 | 0.66 | **0.85** | | | | |
| Continuance Intention (CI) | 0.55 | 0.69 | 0.64 | **0.88** | | | |
| Coercive Pressure (CP) | 0.28 | 0.31 | 0.26 | 0.34 | **0.82** | | |
| Normative Pressure (NP) | 0.33 | 0.37 | 0.29 | 0.36 | 0.49 | **0.83** | |
| Mimetic Pressure (MP) | 0.30 | 0.35 | 0.27 | 0.38 | 0.46 | 0.52 | **0.81** |

Note. Diagonal elements (bold) represent the square root of the average variance extracted (AVE) for each construct. Off-diagonal elements represent inter-construct correlations.

**Appendix I. Structural Model Results**

| **Hypothesis** | **Path** | $\beta$ | **SE** | $z$ | **Result** |
|---|---|---|---|---|---|
| H1 | Confirmation → Perceived Usefulness | 0.63 | 0.05 | 12.60*** | Supported |
| H2 | Confirmation → Satisfaction | 0.32 | 0.06 | 5.33*** | Supported |
| H3 | Perceived Usefulness → Satisfaction | 0.47 | 0.06 | 7.83*** | Supported |
| H4 | Perceived Usefulness → Continuance Intention | 0.41 | 0.07 | 5.86*** | Supported |
| H5 | Satisfaction → Continuance Intention | 0.36 | 0.07 | 5.14*** | Supported |
| H6 | Coercive Pressure → Continuance Intention | 0.12 | 0.05 | 2.40* | Supported |
| H7 | Normative Pressure → Continuance Intention | 0.18 | 0.06 | 3.00** | Supported |
| H8 | Mimetic Pressure → Continuance Intention | 0.21 | 0.06 | 3.50*** | Supported |

Note. Statistical significance is denoted as *** $p < .001$, ** $p < .01$, * $p < .05$.